\documentstyle[twoside,fleqn,espcrc2,epsf]{article}

\newcommand{\ewxy}[2]{\setlength{\epsfxsize}{#2}\epsfbox[10 60 640 570]{#1}}

\let\ga=\gamma

\let\del=\nabla
\let\si=\sigma

\let\om=\omega

\def\to{\rightarrow}

\let\<=\langle
\let\>=\rangle

\let\txt=\textstyle

\def\eqn#1{(\ref{#1})}  

\def\beq{\begin{displaymath}}  
\def\eeq{\end{displaymath}}
\def\bea{\begin{eqnarray*}}
\def\eea{\end{eqnarray*}}

\def\ba{\begin{array}}
\def\ea{\end{array}}

\def\o{\over}

\def\slash{\!\!\!\!/\,}

\def\comment#1{ \hbox{[{\it Comment suppressed here.}\/]} }
\def\hide#1{}

\let\om=\omega

\def\to{\rightarrow}

\let\<=\langle
\let\>=\rangle

\let\txt=\textstyle

\def\eqn#1{(\ref{#1})}  

\def\beq{\begin{equation}}
\def\eeq{\end{equation}}
\def\ba{\begin{array}}
\def\bea{\begin{eqnarray}}
\def\ea{\end{array}}
\def\eea{\end{eqnarray}}

\def\slash{\!\!\!\!/\,}

\def\comment#1{ \hbox{[{\it Comment suppressed here.}\/]} }
\def\hide#1{}

\def\Ord{ {\rm O} }

\def\IR{\relax{\rm I\kern-.18em R}}
\def\IN{\relax{\rm I\kern-.18em N}}
\def\IB{\relax{\rm I\kern-.18em B}}
\def\IE{\relax{\rm I\kern-.18em E}}
\def\ZZ{\relax{\sf Z\kern-.4em Z}}

\def\TT{\mathchoice
       {\sf T\kern-0.52 em{T}}{\sf T\kern-0.52 em{T}}
       {\sf T\kern-0.40 em{T}}{\sf T\kern-0.40 em{T}}}
\def\IP{\mathchoice
       {\sf I\kern-0.14 em{P}}{\sf I\kern-0.14 em{P}}
       {\sf I\kern-0.11 em{P}}{\sf I\kern-0.11 em{P}}}
\def\id{1\kern-.25em {\rm l}}

\newcommand{\skipover}[1]{}
\newcommand{\nn}{\nonumber \\}

\def\half {{\txt {1\over 2}}}

\def\+{\,+\,}
\def\-{\,-\,}

\def\Dssl{{{\bf D}\slash}}

\title{
\vskip -90pt
{\small
\mbox{} \hfill FSU-SCRI-98-89\\
}
\vskip  65pt
	Non-Perturbative 
	Improvement of the Anisotropic Wilson 
	QCD Action\thanks{Work supported by DOE grants 
	DE-FG05-85ER250000 and DE-FG05-96ER40979.}
}

\author{Timothy R. Klassen\address{SCRI, Florida State University, 
			Tallahassee, FL 32306-4130, USA}
        }

\begin{document}       

\begin{abstract}
We describe the first steps in the 
extension of the Symanzik O($a$) improvement program for Wilson-type
quark actions to {\it anisotropic} lattices,
with a temporal lattice spacing  smaller than the spatial one.
This provides a fully relativistic and computationally efficient
framework for the study of heavy quarks.  We illustrate our method
with accurate results for the quenched charmonium spectrum.

\end{abstract}

\kern -10ex
\maketitle


The simulation of charm or bottom quarks on (a sequence of) lattices with
$a m_q \!\ll\! 1$ is currently not affordable, even in the quenched
approximation.  Since standard lattice actions break down when
$a m_q \!>\! 1$, one must design special actions to study heavy
quarks.
Currently two approaches are widely
used: (i) Non-relativistic lattice QCD (NRQCD)~\cite{NRQCD}, 
and ~(ii) the ``heavy relativistic'' or ``Fermilab'' approach~\cite{FNAL}.

The heavy relativistic approach aims to implement the 
improvement program for the Wilson action  at  any quark mass. 
Then     the action only has manifest O(3) symmetry,  and to improve it
to a given order in the spatial momenta requires all coefficients
to be {\it mass-dependent}.
The approach we would like to advocate~\cite{TKSF,TKprep}
can be denoted as ``anisotropic
relativistic'', for short. It is similar to the heavy relativistic one
in that the same terms have to be included in the lattice action, 
but it avoids the mass dependence of the coefficients (except
for one, see below) by using a temporal lattice spacing $a_0$ 
so small  that $a_0 m_q \ll 1$, 
and aiming to improve the action fully only up to O($a$).
It has become clear in the last few years 
that a {\it non-perturbative} determination of the leading
improvement coefficients in a Wilson-type action is very 
important~\cite{ALPHA,EHKprl}.
In our approach this can be achieved, we believe, but in others
the large  number and/or mass-dependence of the 
coefficients 
make this appear very difficult. 



Consider an anisotropic lattice with fixed (renormalized) anisotropy
$\xi = a/a_0$. 
 Up to O($a$) a Wilson-type quark action on such a lattice 
can have the following terms, in continuum notation,
\bea\label{terms}
&& \!\!\! {D\slash}_0 \, (1+\Ord(am))\, , \quad  
		\Dssl \, (1+\Ord(am))\, , \quad\nn[2mm]
&& \!\!\! {a D_0^2}, \,\,{a \, D_k^2}, \,\,
   {a \, [\Dssl,{D\slash}_0]}, \,\,
  a \, \si_{0k} F_{0k}, \,\,
  a \, \si_{kl} F_{kl}  
\eea
Here $\sigma_{\mu\nu}=-{i\over 2} [ \gamma_\mu,\gamma_\nu ]$ and 
$F_{\mu\nu}$ is the field strength.
Using a field transformation we can set the coefficient of ${D\slash}_0$ 
(say) to 1 and adjust the coefficients of the first three terms in the 
second line of~\eqn{terms}. 
 We set the third term to zero, and adjust the
second-order derivatives to combine with the first-order ones 
to give the computationally efficient Wilson operators,
$W_\mu \equiv \del_\mu - \half a_\mu \Delta_\mu$, where $\del_\mu$ and 
$\Delta_\mu$ are the standard 
discretizations of first- and second-order
covariant derivatives
(other possible strategies are discussed 
in~\cite{TKprep}).
This gives a quark matrix $Q$ of the form
\bea\label{genAW}
 Q & \!=\! & m_0 + 
 W_0 \ga_0 +\nu \sum_k W_k \gamma_k \nn[-3mm]
 \quad && - {a\o 2} 
\bigg[{\om_0}  \sum_k \sigma_{0k} F_{0k}
    + {\om}  \sum_{k<l} \sigma_{kl} F_{kl} \bigg] \, .
\eea
Here $\nu$ is a ``bare velocity of light'' and $\om_0$, $\om$ are the
temporal and spatial ``clover coefficients''; 
all three have to be tuned on the quantum level.

\begin{figure}[t]
\vskip -1mm
\centerline{\ewxy{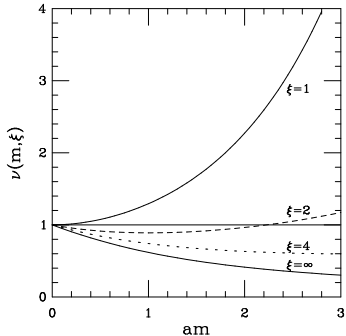}{65mm}}
\vskip -16mm
\caption{Classical bare velocity of light.}
\label{magic_cb_fig}
\vskip -5mm
\end{figure}

\begin{figure}[tbp]
\vskip  0mm
\centerline{\ewxy{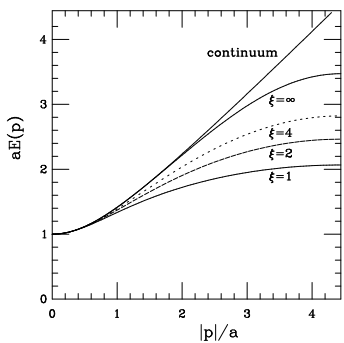}{65mm}}
\vskip -16mm
\caption{Dispersion relation $E({\bf p})$ of a free, massive quark for 
        ${\bf p}\propto (1,1,0)$ after tuning 
        $\nu$. 
        }
\label{DR_magic}
\vskip -5mm
\end{figure}

It is instructive to compare our and the heavy relativistic
approach. The latter corresponds to the special case of an isotropic lattice,
$\xi\!=\!1$, and one must adjust the coefficients $\nu, \om_0, \om$ to all
orders in the mass.
As an example consider the classical  tuning of $\nu$.
Defining an ``effective velocity of light'' c({\bf p}) via
the dispersion relation
 $E({\bf p})^2 = E(0)^2 + c({\bf p})^2 {\bf p}^2$,  
we demand $c(0)\!=\!1$ for  a free quark.     
The result for $\nu$ is shown in figure~\ref{magic_cb_fig} for
various $\xi$, as a function of the pole mass 
$m=\ln(1+a_0 m_0)/a_0$ of the quark.
Note that $\nu$ has a strong mass dependence
in the isotropic case, but only a very weak one for anisotropic lattices.
The dispersion relation 
is shown in 
figure~\ref{DR_magic}.   It is much better for anisotropic lattices,
illustrating that they  are the natural habitat for heavy quarks.

One can also work out the exact classical mass dependence 
of~$\om_0$ and~$\om$. 
{It is~(almost) identical}\linebreak[3] to figure~\ref{magic_cb_fig},
with a strong/weak mass dependence in the isotropic/anisotropic cases. 
In our approach 
we can choose $\om_0$ and $\om$ to be mass-independent, 
but must adjust $\nu$ to O($a$) 
(for details and subtleties see~\cite{TKprep}).
We can also tune $\nu$ 
non-perturbatively, by demanding that the pseudo-scalar meson, say, satisfy 
$c(0)\!=\!1$.
In fact, it is {\it simpler} to determine $\nu$ exactly for the heavy masses of 
interest in this manner; we thereby avoid expensive simulations at small masses 
to determine the O($a^0$) and O($a$) pieces of $\nu(m,\xi)$. 

The complete anisotropic Wilson action 
consists of the quark action described above 
and the anisotropic Wilson 
gauge action (see e.g.~\cite{anisoW}),
which is parameterized by
$\beta$ and the bare anisotropy $\xi_0$. 
For given $\beta$, $\xi$ and $m_0$,  tuning the full action
involves determining 
$\xi_0$, $\nu$, $\om_0$ and $\om$ self-consistently to 
give the same renormalized anisotropy $\xi$ in the gauge and quark sectors,
and no O($a^0,a$) errors. 
 Although not a problem of principle, in full
QCD this is 
a costly tuning problem, so we will here only consider
the {\it quenched} case. Then the relation $\xi_0 =\xi_0(\xi,\beta)$ 
can be determined once and for all
in the pure gauge theory,  which we did to high precision in~\cite{anisoW}. 

To non-perturbatively tune the quark action one can 
use the following  strategy:
(i) Determine $\nu(m,\xi)$ by demanding that the pseudo-scalar 
have a relativistic dispersion relation for $p\to 0$. 
(ii)  Determine the clover coefficients using the Schr\"odinger 
functional with the PCAC relation as improvement condition
 (for details cf.~\cite{TKSF,TKprep,ALPHA,EHKprl}).
 For $\om_0$ impose 
fixed boundary conditions in the time direction to generate a color-electric
background field; similarly, for $\om$ impose fixed boundary conditions
in a spatial direction.

On the classical level this strategy completely decouples
the tuning of the three coefficients.  On the quantum level 
this will not hold exactly, and one will have 
to tune the coefficients iteratively.  Since our $\om_0$ and $\om$ 
can be chosen to be
mass-independent and the mass dependence of $\nu$ is very weak
also on the quantum level for 
$\xi\geq 2$ or so (see below), we do
not expect serious problems.



The above  strategy will require some time and effort to complete. 
Due to the phenomenological importance 
and difficulty of the non-perturbative study of heavy quarks we have
therefore decided to first present a feasibility study of our approach that
does not tune all coefficients non-perturbatively, but only does so for 
the leading one, $\nu$,  and uses a tree-level 
tadpole improvement estimate~\cite{LM} for $\om_0$ 
and $\om$. 
For the tadpole estimate we use the 
mean links in Landau gauge. For the isotropic case it is 
known~\cite{GPLtsukuba}
that this gives an estimate much closer to the non-perturbative 
value~\cite{ALPHA,EHKprl} than the mean plaquette estimate.

We performed simulations of the charmonium spectrum at several couplings
for $\xi\!=\! 3$ and $\xi\!=\! 2$, corresponding to
spatial lattice spacings in the range
$0.17$ to $0.3\,$fm  (the scale was set very accurately 
via $r_0$~\cite{aniso_r0}).
 The bare quark mass $m_0$ and velocity of light $\nu$ 
were  tuned      so that the spin average 1S meson
mass equals its observed value 
and that $c(0)=1$ for the 
pseudo-scalar. $c(0)$ was obtained by extrapolation 
{}from $c({\bf p})$ at the two lowest on-axis momenta.  
We find that the  $\nu$ have a rather weak mass dependence;
they are even quite
close to their classical values of figure~\ref{magic_cb_fig}.
We compared the scale from setting ${r_0}\!=\!0.5\,$fm 
with that from the spin-averaged $1P-1S$ splitting,   
finding that they  agree within a few percent.

\begin{figure}[t]
\vskip  -1mm
\centerline{\hspace{0em}\ewxy{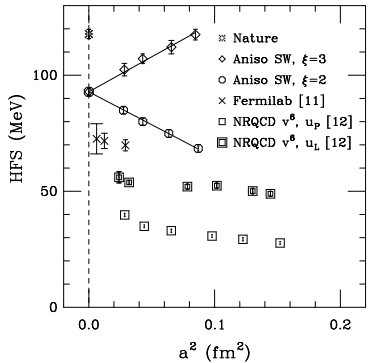}{68mm}}
\vskip  -13mm
\caption{Quenched 
	$c\bar{c}$ hyper-fine splitting $1^3S_1-1^1S_0$ 
        from our ($\diamond$, $\circ$) and other methods.  
	A joint $a^2$ fit of our results is shown. For details
	see~\protect\cite{TKprep}.
        }
\label{HFS}
\vskip  -5mm
\end{figure}

\begin{figure}[tbph]
\vskip  -1mm
\centerline{\hspace{0em}\ewxy{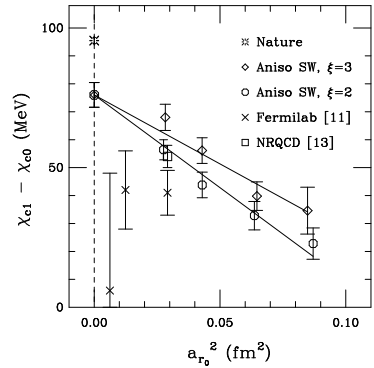}{68mm}}
\vskip  -13mm
\caption{$P$-state splitting $1^3P_1-1^3P_0$.} 
\label{chi_c10}
\vskip  -5mm
\end{figure}

Our main results are the hyper-fine splittings  
for the $S$- and $P$-states shown in 
figures~\ref{HFS} and~\ref{chi_c10}, where they are compared with
results from other approaches. 
Each of our data points uses $400-600$ 
configurations with suitably smeared quark sources for early plateaux.
We regard our results as the most reliable
(note that the strong mass dependence of $\nu$, $\om_0$ and $\om$ is currently
not taken into account in the heavy relativistic approach; also, 
it had been concluded earlier~\cite{Trottier} 
that NRQCD is breaking down for charmonium, at least for spin splittings).
We  emphasize the rather  non-trivial agreement between
the continuum limits of our results for $\xi\!=\!2$ and 3 in an $a^2$ 
continuum extrapolation, which suggests
that our  estimates of $\om_0$, $\om$ 
eliminate most O($a$) errors.

To summarize,
on-shell O($a$) improvement on anisotropic lattices is 
a computationally efficient scheme for the 
study of heavy quarks (all our results were obtained on workstations).
Its efficiency is chiefly due to two factors:
(i) Anisotropic lattices give very accurate effective 
masses. (ii)  Tuning 
$\nu$,  which is the correct way to proceed anyhow, 
is in practice simpler and more accurate than using the ``kinetic
mass prescription'' as in other approaches.
~Probably most important, however, it seems that  systematic errors due
to mass-dependent, relativistic, and quantum corrections 
to the coefficients in
the action can be      more easily
eliminated on anisotropic lattices~\cite{TKprep}.

\end{document}